\documentclass[
 aip,
 amsmath,amssymb,
reprint,%
]{revtex4-2}

\usepackage{graphicx}
\usepackage{dcolumn}
\usepackage{bm}
\usepackage[font=small,labelfont=bf, justification=justified,format=plain]{caption}
\usepackage{adjustbox}
\usepackage{siunitx}

\usepackage[utf8]{inputenc}
\usepackage[T1]{fontenc}
\usepackage{mathptmx}
\usepackage{etoolbox}

\makeatletter
\def\@email#1#2{%
 \endgroup
 \patchcmd{\titleblock@produce}
  {\frontmatter@RRAPformat}
  {\frontmatter@RRAPformat{\produce@RRAP{*#1\href{mailto:#2}{#2}}}\frontmatter@RRAPformat}
  {}{}
}%
\makeatother

\begin{document}

\preprint{AIP/123-QED}

\title{Robust dark-field signal extraction for modulation-based X-ray tensor tomography}

\author{Ginevra Lautizi}
\affiliation{Department of Physics, University of Trieste, Trieste 34127, Italy}
\affiliation{Elettra-Sincrotrone Trieste, Basovizza 34149, Italy}
\email{ginevra.lautizi@phd.units.it}

\author{Vittorio  Di Trapani}
\affiliation{Department of Physics, University of Trieste, Trieste 34127, Italy}
\affiliation{Elettra-Sincrotrone Trieste, Basovizza 34149, Italy}

\author{Alain Studer}
\affiliation{Data Processing Development and Consulting Group, Paul Scherrer Institut, Villigen 5232, Switzerland}

\author{Marie-Christine Zdora}
\affiliation{School of Physics and Astronomy, Monash University, Clayton Campus, Melbourne, VIC 3800, Australia}

\author{Fabio De Marco}
\affiliation{Department of Physics, University of Trieste, Trieste 34127, Italy}
\affiliation{Elettra-Sincrotrone Trieste, Basovizza 34149, Italy}

\author{Jisoo Kim}
\affiliation{Advanced Instrumentation Institute, Korea Research Institute of Standards and Science, Daejeon 34113, Republic of Korea}

\author{Federica Marone}
\affiliation{Photon Science Division, Paul Scherrer Institut, Villigen 5232, Switzerland}

\author{Marco Stampanoni}
\affiliation{Photon Science Division, Paul Scherrer Institut, Villigen 5232, Switzerland}
\affiliation{Institute for Biomedical Engineering, ETH Zürich, Zürich 8092, Switzerland}

\author{Pierre Thibault}
\affiliation{Department of Physics, University of Trieste, Trieste 34127, Italy}
\affiliation{Elettra-Sincrotrone Trieste, Basovizza 34149, Italy}

\date{\today}

\begin{abstract}
We demonstrate a robust signal extraction method for X-ray speckle-based tensor tomography. We validate the effectiveness of the method for several carbon fiber composites, highlighting its potential for industrial applications. The proposed method can be adapted to various acquisition schemes and wavefront-marking optical elements, making it a versatile and robust tool for X-ray scattering tensor tomography.
\end{abstract}

\maketitle

Modulation-based imaging (MBI) exploits heterogeneous illumination patterns to locally encode the X-ray interaction with a sample. While a uniform illumination can normally reveal only the attenuating properties of the sample, a non-uniform X-ray illumination can be used to encode the otherwise invisible X-ray refraction and ultra-small-angle scattering effects in the sample.\cite{Pfeiffer2008, Olivo2021} This concept is most commonly implemented with regular grid patterns, such as in (Talbot-)Lau interferometers. \cite{David2002, Momose2003} Encoding and decoding of these signals can also be achieved with other types of illumination profile. In particular, speckle-based imaging (SBI) uses a diffuser, for instance one or more layers of sandpaper, to allow for extraction of the three complementary image signals: attenuation, refraction, and small-angle scattering, also known as dark-field signal.\cite{Morgan2012, Berujon2012, Zanette2015, Zdora2018, Zhou2018, Zdora2021, DeMarco2023}

The X-ray dark-field signal is a manifestation of the small-angle scattering properties of a sample, which are caused by inhomogeneities in the local electron density at a scale smaller than the imaging system's resolution.\cite{Pfeiffer2008} Samples with strongly oriented microstructural features often exhibit an anisotropic X-ray scattering profile. In these cases, the directionality of the scattering can also be extracted to reveal information about their orientation, e.g., of fibers in a composite material. 
This directional information is conventionally measured in small-angle X-ray scattering (SAXS) tomography experiments,\cite{Schaff2015, Liebi2015} and maps can be obtained by scanning a pencil beam on the sample.\cite{Jensen2010} It is now known that MBI methods can also provide directional information through a careful analysis of the pattern distortions.\cite{Wang2015, Kagias2016, Pavlov2021, Smith2022}

When measured for various orientations of the sample, two-dimensional directional scattering signals can be combined to tomographically reconstruct the local scattering tensor of the sample using X-ray tensor tomography (XTT).\cite{Malecki2014, Vogel2015, Schaff2015, Sharma2016, Sharma2017, Felsner2019, Kim2020}Compared to traditional X-ray microtomography, XTT enables the study of the microstructural organization in significantly larger volumes without the need for high spatial resolution of the imaging system.\cite{Kagias2019, Kim2021, Kim2022} As a new microtomography modality, XTT has great potential for applications in different areas of research, such as the detection of wrinkling and fiber-waviness within fiber-reinforced polymers (FRPs) and the investigation of fibrous tissues in biological samples.

The fast full-field modality of MBI methods makes it an excellent contender for XTT applications\cite{Lautizi2024, Kim2021}. X-ray speckle-based directional dark-field imaging was first introduced by Wang et al.\cite{Wang2015}, but required precise diffuser movements and hundreds of scanning points, making it time-consuming and prone to artifacts. Zhou et al.\cite{Zhou2018} addressed these limitations by developing a method requiring only two images, enabling efficient reconstruction with less stringent setups. Pavlov et al.\cite{Pavlov2021} later proposed a Fokker-Planck-based approach for direct extraction of directional signals, but assumed slow signal variation, limiting its applicability to cases where rapid changes in scattering properties occur. Recently, Smith et al.\cite{Smith2022} managed to extract detailed scattering properties, but faced slow convergence, hindering real-time tensor tomography applications.

In this Letter, we present a robust method for the extraction of the directional dark-field signal from data acquired with random diffusers. 
For a random diffuser, the signal extraction method described in Ref. \cite{Lautizi2024} struggles when calculating the 2D scattering tensor in some regions for numerical instabilities caused by abrupt local variations in intensity.
The main difference of our omnidirectional dark-field signal extraction method compared to earlier studies lies in the robustness and the versatility of our full-field method, which is applicable to various wavefront modulators, including non-periodic ones.
We apply this approach to experimental measurements and demonstrate the first reconstruction of X-ray speckle-based tensor tomography.

\begin{figure}[h!]
    \centering
    \includegraphics[width=0.8\linewidth]{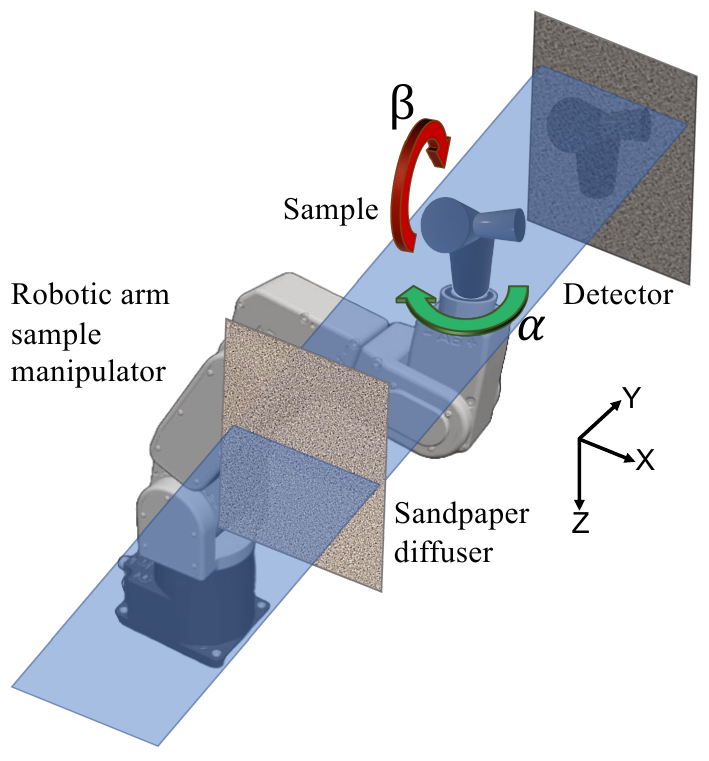}
    \caption{Schematic overview of the experimental setup. The X-ray beam is modulated by a diffuser mounted on a scanning stage upstream of the robotic arm used as the sample manipulator. The detector system consists of an sCMOS camera with an X-ray microscope.}
    \label{fig:fig0}
\end{figure}

A full MBI-XTT dataset is made of multiple image acquisitions that encode complementary scattering information through diverse sample positioning. A schematic of our experiment is shown in Fig.~\ref{fig:fig0}. 

Unlike conventional tomography, the dataset must contain multiple full tomographic scans (where the angle $\alpha$ is sampled over the full $\SI{360}{\degree}$ range) taken at different tilts of the axis of rotation $\beta$. For each pair $(\alpha, \beta)$, multiple lateral displacements of the sample with respect to the patterned illumination are recorded, as is normally done with MBI methods.

The results presented in this letter were obtained from an experiment conducted at the SYRMEP beamline (Elettra Sincrotrone Trieste) using a filtered white beam with a mean energy of $\SI{37.6}{\kilo\eV}$.
Using a filtered white beam, instead of a monochromatic beam, supports the translation to a laboratory-based setup.

To get lateral diversity, we employed the diffuser-stepping approach of SBI, which involves translating the diffuser to $20$ different transverse positions while keeping the sample fixed \cite{DeMarco2023}.

We used three layers of P320 sandpaper as a diffuser. The detector was a water-cooled sCMOS camera (Orca Flash 4.0, Hamamatsu) coupled with an X-ray microscope (Optique Peter), with an effective pixel size of $\SI{3.82}{\micro\metre}$. The illuminated region of the sensor was $1452 \times 2048$ pixels, resulting in a field of view of $5.5 \times 7.8 \SI{}{mm^2}$. In our experiment, the sample-detector distance was $\SI{52}{\cm}$, while the diffuser-sample distance was  $\SI{92.5}{\cm}$. The exposure time for each projection was $\SI{75}{\milli\second}$. The sample manipulator was a Meca500 robotic arm (Mecademic Robotics, Montreal, Canada) with a repeatability of $\SI{5}{\micro\metre}$ and six degrees of freedom. \cite{DiTrapani2024}

Extraction of the directional scattering information is done on pairs of images (with sample / without sample). The analysis is performed over a local window within the image, which is translated to cover all pixels. For each local window, the scattering signal can be modeled, in first approximation, as an anisotropic Gaussian function. 
Therefore, for a single projection at the $j$-th diffuser position and for each local window, we can model the sample intensity pattern in Fourier space as follows:
\begin{equation}
    I_s^{j}(\mathrm{k}) = e^{-f(\mathrm{k})}I_0^{j}(\mathrm{k}).
\label{eq:proj_model}
\end{equation}
In Eq.~\ref{eq:proj_model} $I_s^{j}(\mathrm{k})$ and $I_0^{j}(\mathrm{k})$ are the magnitudes of the Fourier transforms of the window pairs, $\mathrm{k}$ is the Fourier space coordinate, and $f(\mathrm{k})$ models the local attenuation and Gaussian scattering function $f(\mathrm{k}) = \mu + \frac{1}{2}(a k_x^2 + b k_y^2 + c k_x k_y)$.

With speckle measurements, the trivial approach using $f(\mathrm{k}) = - \ln (I_s/I_0)$ is numerically unstable. It is therefore necessary to combine information from all the measured frames obtained for multiple diffuser positions.

The numerical method we introduce here solves the following non-linear least-squares problem:
\begin{equation}
    \mathcal{L} = \sum_{j,\mathrm{k}} w_{\mathrm{k}}^{j} |I_s^{j}(\mathrm{k}) - e^{-f(\mathrm{k})}I_0^{j}(\mathrm{k})|^2,
    \label{eq:L_1}
\end{equation}
with respect to the parameters of $f$. The statistical weights $w_{\mathrm{k}}^{j}$ can be used to account for known uncertainties. Since $I_s^j$ and $I_0^j$ are discrete Fourier transforms of input data windows, these weights could be set proportional to the detector's modulation transfer function, if it is known. 

Close to the optimum, $\mathcal{L}$ in Eq.~\ref{eq:L_1} can be linearized using logarithms, resulting in 

\begin{equation}
    \mathcal{L} = \sum_{j,\mathrm{k}} \mathrm{W}_{\mathrm{k}}^{j} \bigg{|}f - \ln \bigg{(}\frac{I_0^{j}(\mathrm{k})}{I_s^{j}(\mathrm{k})}\bigg{)}\bigg{|}^2 ,
    \label{eq:L_lin}
\end{equation}
where we labelled $w_{\mathrm{k}}^{j}|I_s^{j}(\mathrm{k})|^2$ with $W_{\mathrm{k}}$. This is now a weighted linear least-squares problem of the form $\mathcal{L} = |\mathrm{A}x - b|^2$, which can be solved readily using exist?.ing routines. It is worth noting that if $I_s^{j}(\mathrm{k})$ is close to $0$, the numerical stability is assured by the weight $W_{\mathrm{k}} \propto |I_s^{j}(\mathrm{k})|^2$.
We obtain the 2D scattering tensor for each local window by solving the weighted linear least-squares problem in Eq.~\eqref{eq:L_lin}. The 2D scattering tensor can then be eigendecomposed to obtain eigenvalues and eigenvectors. The mean of the eigenvalues represents the mean scattering signal, the fractional anisotropy \cite{Basser1996} describes how well-aligned the fibers are in each voxel, and the eigenvector with the shortest length is associated with the preferential local fiber orientation.


To validate the method, we acquired SBI projections of four different samples fabricated with carbon fiber materials. In all cases, we used an analysis window of $6\times6$ pixels.
This window is a good compromise between a high spatial resolution and a low noise level in the images. The computation time also plays a role in the decision of a proper analysis window.

\begin{figure}[h!]
    \includegraphics[width=\linewidth]{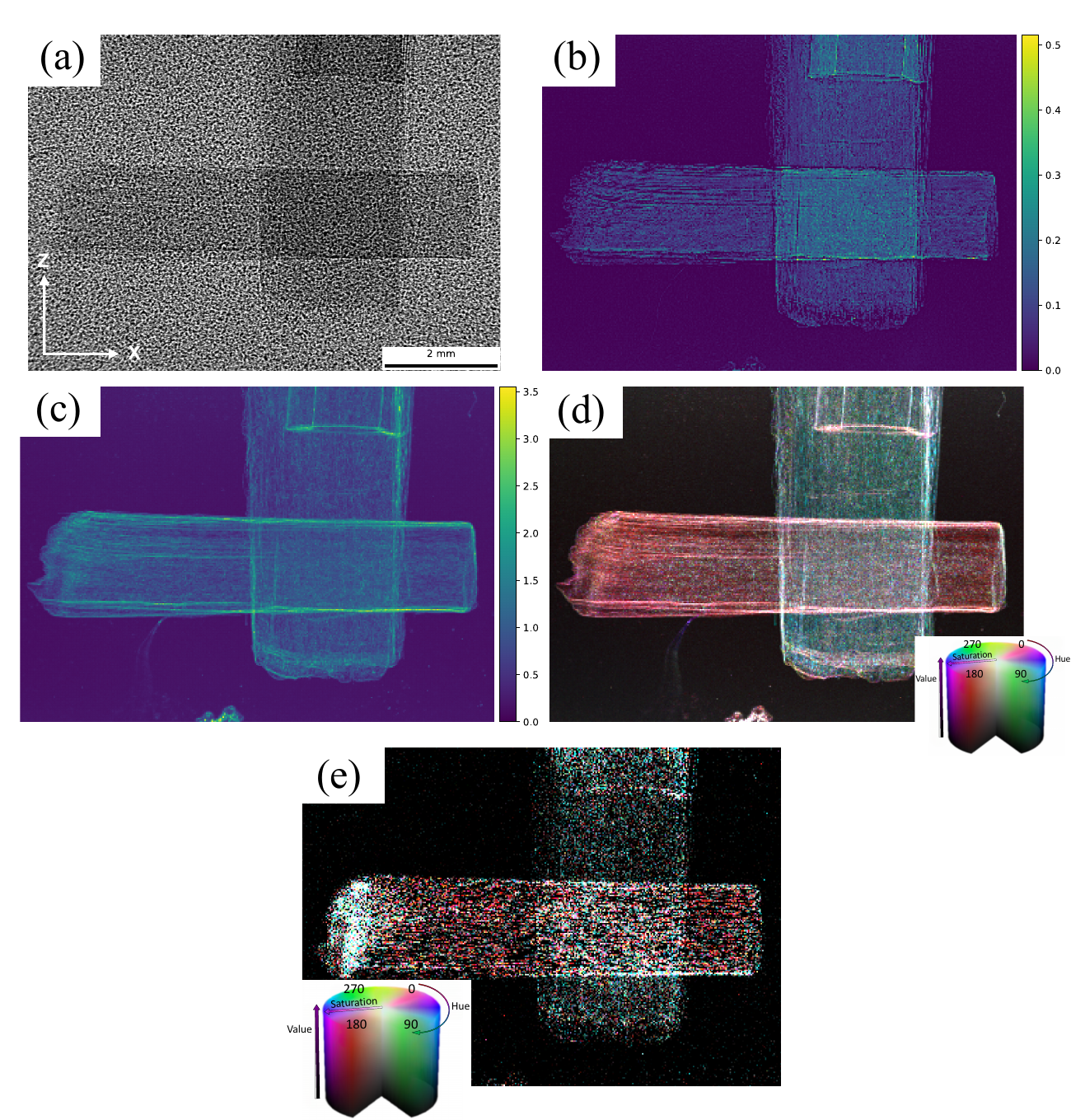}
    \caption{(a) One of the raw transmission images of the carbon fiber cross sample, including the speckle pattern. (b) Extracted absorption image and (c) mean scattering signal in arbitrary units. (d) Composite image in which the orientation is mapped to the hue, scattering anisotropy to saturation, and scattering strength to intensity. As expected, the main orientation signals are seen to be along the horizontal and vertical directions (red and cyan). (e) Composite image of the directional dark-field signal extracted with the method presented in ref.~\citenum{Lautizi2024}. Contrast has been enhanced for visualization purposes by linear scaling.}
    \label{fig:fig1}
\end{figure}

The first test sample was made of two sheets of unidirectional carbon fibers, with a mean diameter between $\SI{5}{\micro\metre}$ and $\SI{10}{\micro\metre}$, glued together at an angle of $\SI{90}{\degree}$. This sample was already analyzed with our previous reconstruction method.\cite{Lautizi2024} The result of the 2D scattering analysis is shown in Fig.~\ref{fig:fig1}. Figure \ref{fig:fig1}a shows one of the $20$ frames that form the complete dataset. The extracted absorption and scattering strength are displayed in \ref{fig:fig1}b--c. In the latter, we see that beyond the strong scattering by the sharp edges, a clear scattering signal is also recorded within the sample itself. These two images correspond to the conventional, non-directional analysis generally obtained with SBI. Fig.~\ref{fig:fig1}d shows the full scattering tensor parameters using an HSV color map. Comparison with our previous approach (Fig.~\ref{fig:fig1}e) shows the improved robustness of the new extraction method.

?.

\begin{figure}[h!]
    \centering
    \includegraphics[width=\linewidth]{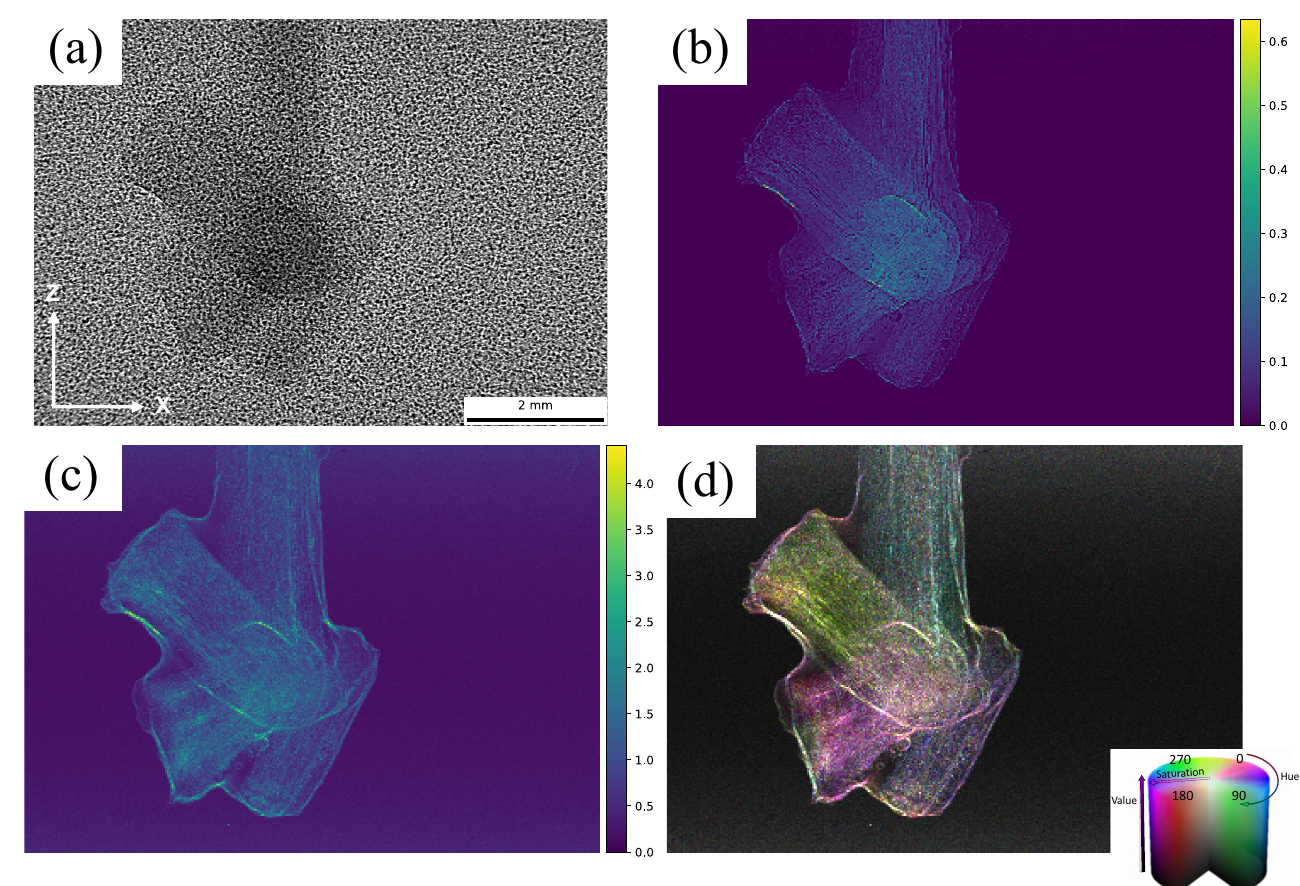}
    \caption{(a) One of the raw transmission images of a sample made of four carbon-fiber rods. (b) Extracted absorption image and (c) mean scattering signal in arbitrary units. (d) Composite image showing the main orientation signals. Contrast has been enhanced for visualization purposes by linear scaling.}
    \label{fig:fig2}
\end{figure}

The results of the 2D scattering analysis on a second test sample are shown in Fig.~\ref{fig:fig2}. This sample was made using four rods cut and glued in four different positions. Each rod was made by unidirectional carbon fibers oriented along the rod lengths. The main-orientation signal confirms that the fibers run along the long axis of the rods.

\begin{figure}[h!]
    \centering
    \includegraphics[width=\linewidth]{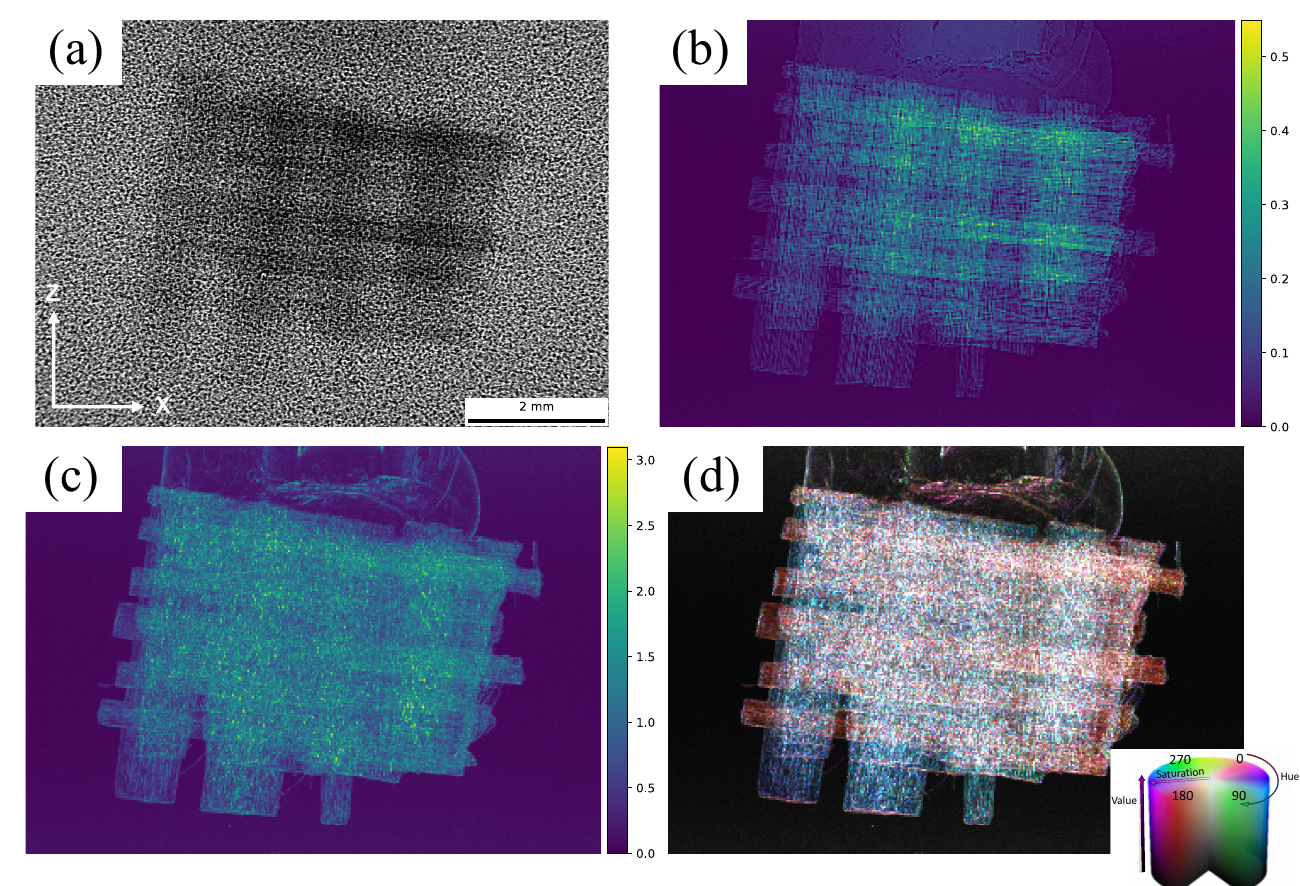}
    \caption{(a) One of the raw transmission images of two glued glass-fiber sheets. (b) Extracted absorption image and (c) mean scattering signal in arbitrary units. (d) Main orientation signals (cyan and red). Contrast has been enhanced for visualization purposes by linear scaling.}
    \label{fig:fig3}
\end{figure}

The third test sample (Fig.~\ref{fig:fig3}) was made of two sheets of unidirectional glass-fiber tape ($\SI{15}{\micro\metre}$ diameter), glued together at an angle of $\SI{90}{\degree}$. Also for this sample, the main-orientation signal confirms that the fibers run along the long axis of the sample. In the regions where the fibers cross at $\SI{90}{\degree}$ the low saturation of the colors indicates a low anisotropy value, as expected for a region where multiple orientations overlap.

Following the demonstration of the 2D orientation analysis, we show that our proposed algorithm enables X-ray speckle-based tensor tomography. For this demonstration, we performed an XTT measurement of the carbon-fiber sample shown in Fig.~\ref{fig:fig2}. 
We used the stair-wise acquisition protocol described in ref. \citenum{Kim2020} and \citenum{Kim2021}, varying the tilt angle $\beta$ from $\SI{0}{\degree}$ to $\SI{40}{\degree}$, with an angular step of $\SI{10}{\degree}$. For each angle $\beta$, we continuously rotated the sample over $\SI{360}{\degree}$, acquiring $360$ projections with the setup shown in Fig.~\ref{fig:fig0}. This procedure was repeated for $20$ different positions of the diffuser.
The omnidirectional dark-field signal was then extracted from all projections of the tomographic dataset, thus providing the 2D scattering tensor for each local window and for each projection. These results are combined to make the tensor sinogram describing the full scattering tensor field in 3D.\cite{Lautizi2024} For this reconstruction, we used an analysis window of $6\times6$ pixels. The projections were aligned using a customized version of the alignment algorithm described in ref. \citenum{Liebi2015}.

\begin{figure}[h!]
    \centering
    \includegraphics[width=\linewidth]{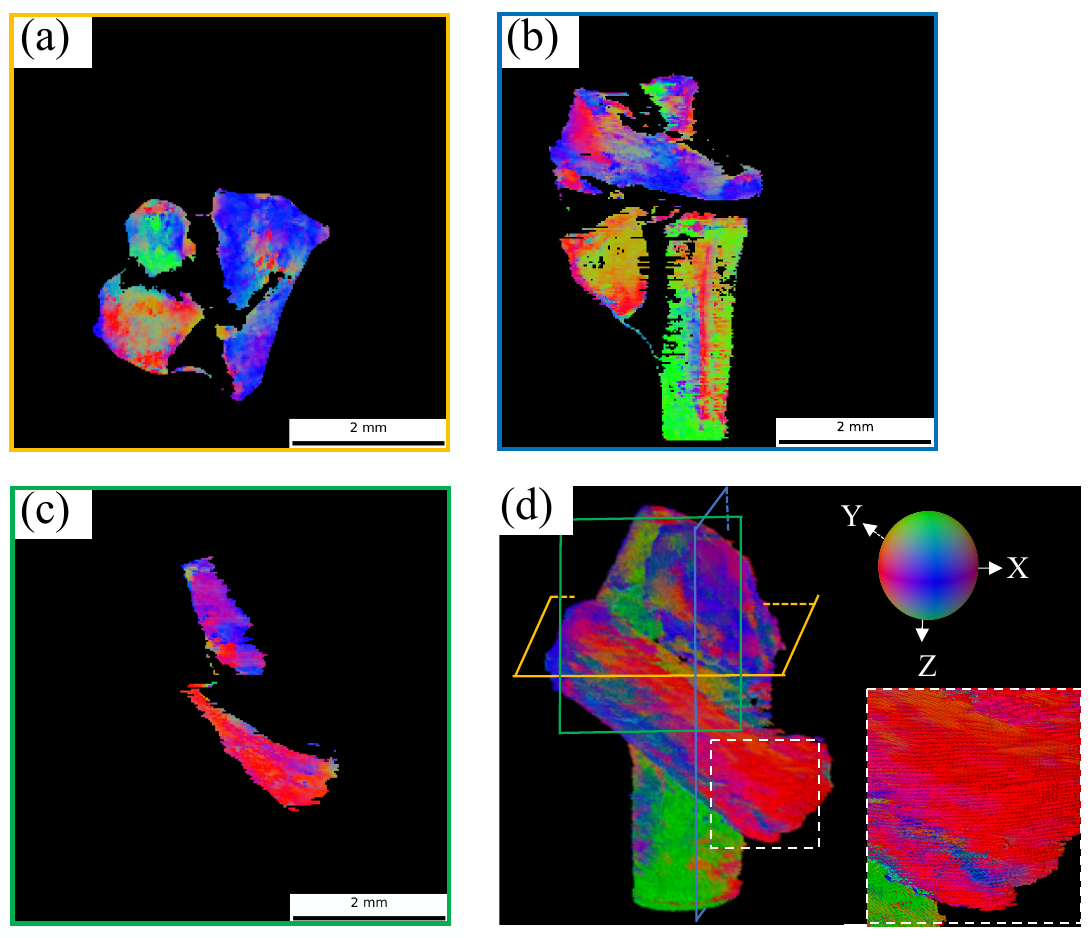}
    \caption{XTT results of a carbon fiber sample made using $4$ unidirectional carbon fibers rods cut and glued in $4$ different positions. (a) An axial slice, (b) a coronal slice and (c) a sagittal slice of the sample volume. The color is an RGB representation of the local structure orientation. The color ball is shown in (d) and it is symmetric with respect to the x - y, x - z, and y - z planes. (d) A 3D visualization of the reconstructed scattering tensor. In this representation, each arrow's orientation corresponds to the main direction in each voxel. The colorful planes indicate the slices shown in the other panels. The white inset shows a zoomed region to appreciate the directional information}.
    \label{fig:fig5}
\end{figure}

Tomographic slices of the tensor tomogram are displayed in Fig.~\ref{fig:fig5}. 
Three slices of the sample volume are shown: an axial slice (Fig.~\ref{fig:fig5}a), a coronal slice (Fig.~\ref{fig:fig5}b), and a sagittal slice (Fig.~\ref{fig:fig5}c). Since the local structure-orientation signal is prone to noise, in particular in background regions and at sharp edges, the tomographic volume was masked with a threshold-based mask that was both based on the average scattering and the absorption signals. The glue used to fix the four parts of the sample contributed to the scattering signal and was therefore segmented out.

In the tomographic slices, the color represents the local structure orientation of the fibers within each voxel, where the voxel size is given by the size of the analysis window. A 3D visualization of the scattering tensor reconstruction of the sample is shown in Fig.~\ref{fig:fig5}d. Each arrow's direction and color (RGB) represent the main orientation direction within a voxel. 

As visible both in the tomographic slices and in the 3D visualization, the four main orientations of the carbon fiber rods were extracted. At the edges, the colors are not perfectly uniform, which we attribute to damage to the rods during sample preparation and to unsegmented glue.

In conclusion, we have presented a demonstration of speckle-based tensor tomography. This achievement was made possible by developing a method to optimally extract the X-ray directional dark-field signal for data acquired with a random diffuser.
Formulating the problem with a weighted linear least-squares approach guaranteed the numerical stability needed for a robust reconstruction.
We demonstrated the effectiveness of our approach for SBI projection scans of different samples and compared the results with previous findings on the same sample. The new algorithm proved to be more stable, resulting in less noisy dark-field images than our previous approach.

The proposed method, applicable to different data acquisition schemes, creates new opportunities for the applications of X-ray scattering tensor tomography using a non-periodic X-ray beam modulator. 
Our approach operates independently of the experimental geometries, reconstructing the full tensor field.

Thanks to its experimental simplicity, SBI has been demonstrated using laboratory setups, \cite{Zanette2014, Zhou2015} hence speckle-based tensor tomography could also be performed using laboratory sources, thus reaching a broader community.

In the near future, we plan to modify the acquisition procedures testing a sample-scanning approach \cite{DeMarco2023} rather than diffuser-stepping to increase the field of view, minimizing the acquisition time. 

A current limitation of the model is that for each analysis window  of a projection, a single orientation is extracted. Therefore, ongoing research is aimed at ensuring reliable reconstruction even in the case where multiple scattering orientations overlap in a single analysis window, which could significantly enhance the versatility and robustness of this approach.

The data and the code that support the findings of this study are available from the corresponding author upon reasonable request.

Authors wish to thank Elettra Sincrotrone Trieste  for the provision of the beamtime 20215810, and Dr. Giuliana Tromba and Dr. Adriano Contillo for assistance in using the SYRMEP beamline.
This publication is part of a project that has received funding from the European Research Council (ERC) under the European Union’s Horizon 2020 research and innovation program (Grant agreement No. 866026).

\nocite{*}
\clearpage
\bibliography{aipsamp}

\end{document}